\documentclass[18pt,preprint]{aastex}
\begin{document}
\title{Metallicity determination in gas-rich galaxies with semiempirical methods. }
\author{A.M. Hidalgo-G\'amez}
\affil{Department of Physics, Escuela Superior de F\'{\i}sica y Matem\'aticas, IPN, U.P. Adolfo L\'opez Mateos, 
C.P. 07738, Mexico city, Mexico}
\author{ and D. Ram\'{\i}rez-Fuentes}
\affil{Instituto de Astronom\'\i a, UNAM, Ciudad Universitaria, Aptdo. 70 264,
C.P. 04510, Mexico City, Mexico}

\begin{abstract}

A study of the precision of the semiempirical methods used in the determination of the 
chemical abundances 
in gas-rich galaxies is carried out. In order to do this  the oxygen abundances of a total 
of 438 galaxies were determined using the electronic temperature, the $R_{23}$ and the P 
methods. The new calibration of the P method gives the smaller dispersion for the low 
and high metallicity regions, while the best numbers in the turnaround region are given 
by the $R_{23}$ method. We also found that the dispersion correlates with the metallicity. 
Finally, it can be said that all the semiempirical methods studied here are quite 
insensitive to metallicity with a value of $8.0\pm0.2$ dex for more than $50\%$ of the total sample. 


\end{abstract}

\section{Introduction}

The determination of nebular abundances in H\,{\sc ii} regions is not a simple matter. 
It is well known that the ionic abundances depend on the intensity of the lines involved, 
and on the electronic density and temperature (Peimbert \& Torres-Peimbert 1977; Aller 
1984; Osterbrock 1989). For nearby bright galaxies, the weak auroral lines needed for 
the electronic temperature ($T_e$) determination can be detected when the signal-to-noise 
(S/N) is high enough. When the $T_e$ is determined, the ionic abundances can be easily 
obtained through the emissivities of the specific ions, when a certain ionization 
structure for the nebulae is assumed (e.g., Aller 1984;  Izotov et al. 2006 for a more
recent review). Such a procedure is normally called the ``standard method'' for the 
abundance determination. The most important caveat concerning the standard method is 
that the auroral lines are always very weak. The typical uncertainty associated with 
the standard method is about $0.1$ dex (Kewley \& Ellison 2008) but this increase significantly 
for low S/N spectra. 

The situation gets worse when the forbidden auroral lines are absent. The most common
reason for such absence is the low S/N of the spectra but it is not the only one (see
Hoyos \& D\'{\i}az 2006). In such situations the so-called semiempirical, or bright-line, 
methods need to be used in order to determine the metal content of the H\,{\sc ii} region.
The most common and popular of these methods is the $R_{23}$ calibrator proposed by Pagel 
et al. (1979) and improved by McGaugh (1991), among others. Other calibrations of the 
$R_{23}$ method have been proposed using larger samples and more complete stellar 
evolutionary grids (e.g. Zaristky et al. 1994; Kewley \& Dopita (2002); Kobulnicky \& 
Kewley 2004). We prefer to use the calibration by McGaugh (1991). The main reason 
is that, although this calibration is old, it takes into account the influence of 
the ionization parameter on the chemical abundance determination.

In recent years, many other  semiempirical methods have been proposed, based on the 
same lines as the $P$ method (Pilyugin 2000,2001; Pilyugin \& Thuan 2005), or on the 
nitrogen (the $N2$, Denicol\'o et al. 2002) or the sulfur lines (the  S$_{23}$, D\'{\i}az 
\& P\'erez-Montero 2000; the $S_{234}$, Oey \& Shields 2000)

All of the semiempirical methods present problems: bivaluation, large dispersion, dependence 
on other parameters (such as the ionization parameter or the nitrogen abundance), very large 
wavelength range between the lines involved, etc. Moreover, the large dispersion in the 
metallicity values obtained with the semiempirical methods can be due to the different 
H\,{\sc ii} regions geometries (which amount to different ionization parameters) and 
differences in age of the H\,{\sc ii} regions used for the calibration of the semiempirical 
methods. In addition, there are other influences such as the different apertures for different 
spectra which might mask the real calibration of the semiempirical methods (A.M. Hidalgo-G\'amez  
2009, in preparation).

The main interest of this investigation is to determine whether any of the semiempirical methods
available or any of their different calibrations give better oxygen abundances than the 
others. In order to make such a comparison we think that the standard method abundances are 
good touchstones. The idea for this investigation comes out from to the neccesity of realiable abundance 
determinations with any of the semiempirical methods for a sample of dwarf spiral galaxies 
where the [OIII] $\lambda$4363 is not detected (A.M. Hidalgo-G\'amez et al 2009, in preparation). It is well known that galaxies with and without 
the [O\,{\sc iii}]$\lambda$4363\AA~ line have different properties on the $T_e$ range, 
ionization parameter, and luminosity (Hoyos \& D\'{\i}az 2006). Therefore, the metal 
contents might significantly differ. Although we are aware of this, studies such as the one 
presented here might give some clues about the goodness of each of the semiempirical 
methods.

In the next section, the description of the sample of galaxies used in this investigation is 
carried out, while the comparison of the standard method with the semiempirical methods 
used is presented in Section 3. A comparison with other studies already published is presented 
in Section 4, and a brief discussion is given in Section 5. Finally, conclusions are presented 
in Section 6.

\section{Description of the data and the methods}

A homogeneous sample is needed in order to make a proper study of the difference in the 
chemical abundances provided by different methods. By homogeneous we mean a sample observed, 
reduced, and analyzed in the same way in order to reduce the dispersion of the results. 
Moreover, in order to consider the effects of the age of the H\,{\sc ii} regions, the galaxies 
should be of similar morphological type. However, most of the studies done so far on the 
calibrations of the semiempirical methods use samples of galaxies where the oxygen abundances 
have been collected from the literature. Such a procedure might not be very appropriated,
leading to results that are not very reliable. 

In order to take into account all of the possible sources of dispersion, we used a sample as 
homogeneous as possible and also large enough to achieve conclusive results. The sample 
consists of a total of 438 galaxies from Kniazev et al. (2004), mainly blue compact (392) 
and irregular galaxies (28). They were selected from a sample of 612 galaxies observed 
by Sloan Digital Sky Survey (SDSS) where the oxygen line [OIII] $\lambda$4363 is detected. Therefore, the electronic 
temperature can be determined and the standard method can be used for the determination 
of the chemical abundances for all the galaxies. We chose only those galaxies with 
[OII] $\lambda$3727 detected in order to decrease the uncertainties. 

The main caveat of this sample is that, any sample selected by the presence (or absence) 
of a particular spectral line is biased in itself. This should be taken into account when
appliying any calibration based on a sample of galaxies with the [O\,{\sc iii}]$\lambda$4363\AA~ 
line to objects where this line is absent or to a sample with different systematics than
those presented here. In particular, those  galaxies with the presence of the 
[O\,{\sc iii}]$\lambda$4363\AA~ line seem to show higher ionization parameters and less
evolved stellar population than those galaxies without it (Hoyos \& D\'{\i}az 2006). In 
addition, the equivalent width (EW) of H$\beta$ is much larger (a factor of $2.8$) for
the galaxies in the first group than those galaxies without the 
[O\,{\sc iii}]$\lambda$4363\AA~ line.

\begin{figure}
\centering
\includegraphics[width=10cm]{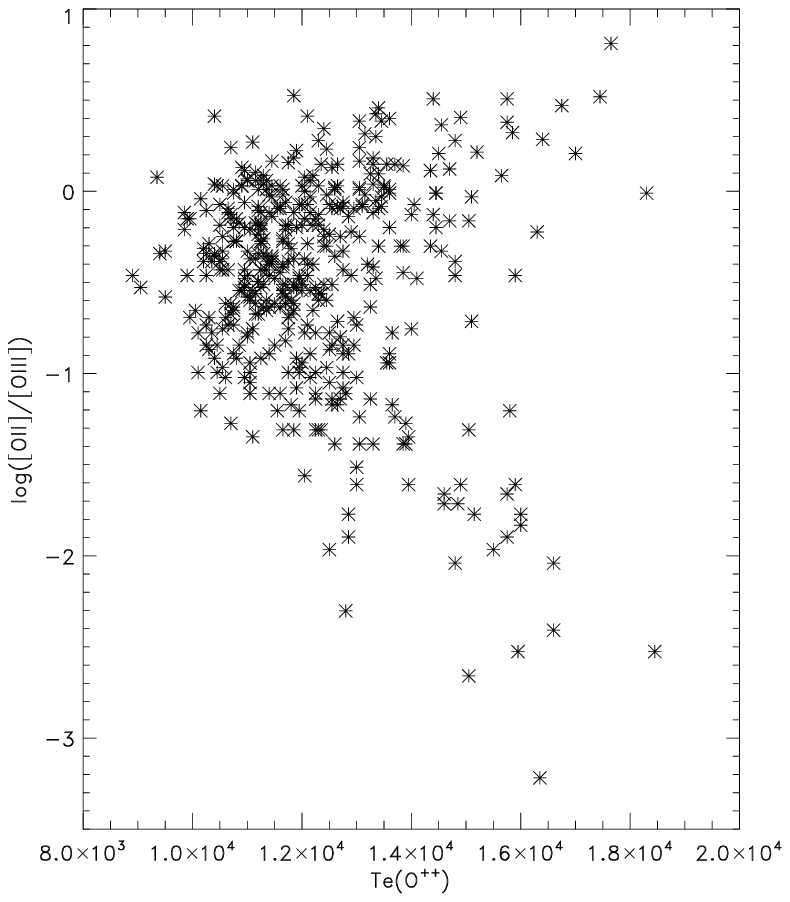}
\caption{Electron temperature vs. the log[O\,{\sc ii}]/[O\,{\sc iii}] ratio. 
The bulk of galaxies in the sample, with temperature lower than $14,000$ K, do 
not show any relationship with the ionization parameter. For those galaxies with 
$T_e$ higher than $14,000$ K there are two different trends: some of them have 
low values of the log[O\,{\sc ii}]/[O\,{\sc iii}] ratio while some others show 
values of this ratio about $1$. }
\label{fig1}
\end{figure}

One of the most interesting parameters in the characterization of the H\,{\sc  ii} regions
is the ionization parameter. It depends on the ionizing continuum and on the geometry of the
region. The ionization parameter is defined in terms of a uniform density, $n_e$, by 
$U=L/4~ \pi~ R_s^2~c$, where $R_s^2$ is the Str\"omgrem radius and $L$ is the H-ionizing 
photon luminosity (Mathis 2000). As we will discuss later, some of the semiempirical 
methods depend on this parameter. Therefore, it will be very interesting to study the 
distribution of the ionization parameter in our sample of galaxies. A simple way of  
doing this is through the ratio [O\,{\sc ii}]/[O\,{\sc iii}]. More than half of the galaxies 
in our sample have values of this ratio between $0.5$ and $1$: only $6\%$ show ratios 
larger than $1.2$ and only $3\%$ smaller than $0.1$. These values are slightly smaller than 
those of the [O\,{\sc iii}]$\lambda$4363 galaxies in Hoyos \& D\'{\i}az (2006). One 
reason for these differences might be that here we have considered only one of the 
[O\,{\sc iii}] lines (at $5007$\AA). We can also check if the $T_e$ (and, therefore, 
the abundances) depends on the ionization parameter. Such a relationship is clear for 
a small sample of H\,{\sc ii} regions in nearby dwarf irregular galaxies (A.M. 
Hidalgo-G\'amez, unpublished results). It is apparent from Figure ~\ref{fig1} that there
is no correlation between $T_e$ and the log([O\,{\sc ii}]/[O\,{\sc iii}]) for those galaxies
with temperatures smaller than  14,000 K and therefore, between the  oxygen content and 
the ionization parameter. However, for those galaxies with $T_e$ higher than 14,000 K, 
there is a group of galaxies with a negative trend (very low values of log([O\,{\sc ii}]/[O\,{\sc iii}])) 
and another with a positive one (log([O\,{\sc ii}]/[O\,{\sc iii}]) $\approx 0.5$). In 
any case, the number of those galaxies is very small (less than $10\%$) to have a great 
influence on our results. 

\subsection{On the standard method}

The standard method (SM) abundance were determined by 
us from the extinction (and absorption) corrected line intensities given by Kniazev 
et al. (2004) with a five-level atom two-zone model (N. Bergvall 1999, private 
communication). A density of $100$ cm$^{-1}$ was considered for all the spectra. 
The electronic temperature of the O$^{++}$ is determined with the ratio between 
the nebular and the auroral oxygen lines, while the temperature of the O$^+$ zone 
is based on photoionization models. From these values a parametrization can be 
obtained as 

\begin{equation}
 T(O^{+}) = 6520 + 0.50 T(O^{++}). 
\end{equation}
The average uncertainty of this equation is about $300$ K. 

In order to see if the parametrization gives the correct values for the $T(O^{++}$) 
a comparison with other procedures can be done.

The electronic temperature of the O$^{+}$ can be determined from the ratio of the 
intensities of the [O\,{\sc ii}] lines at $\lambda$,3726,3739 and $\lambda$,7319,7320.
We found two different equations that related the $T(O^{+})$ and this ratio (hereafter, 
RO2). 
\begin{equation}
T(O^{+}) = {0.853 / log~(I3727/I7325) - 0.928 + log~(1~+~7.03~x) + 0.02~log~t}
\end{equation}
 
where $I3727$ is the intensities of the lines [O\,{\sc ii}]$\lambda$,$\lambda$3726,3729, 
$I7325$ the intensities of the lines [O\,{\sc ii}]$\lambda$,$\lambda$7320,7330, $t$
is the $T(O^{++})$ in units of $10^4$, and $x = 0.01 N_e/T_e$ where $N_e$ is the 
electronic density (H\"agele et al. 2008). The other equation is

\begin{equation}
T(O^{+}) = a_o + a_1~RO2 + {a_2/ RO2}
\end{equation}

with $a_o = 0.23 - 0.0005~n -{0.17/n}$, $a_1 = 0.0017 + 9~10^{-6}~n + {0.0064/n}$, and  
$a_2 = 38.3 - 0.021~n - {16.4/n}$, where $n$ is the density (H\"agele et al. 2008). 

When the auroral lines cannot 
be detected because of the low S/N or the small wavelength range of the spectra, the  
$T(O^{+})$ can be determined from the parametrization between $T(O^{+})$ and $T(O^{++})$
using photoionization models, such as, e.g., Campbel et al. (1986), or using observational data such 
as Pilyugin et al. (2006). In order to see how accurate our equation is, we have compared
the $T(O^{+})$ determined from it with the values obtained with the parametrizations by 
Pagel et al. (1992), Izotov et al. (2006), Deharveng et al. (2000), Pilyugin et al. 
(2006), P\'erez-Montero \& D\'{\i}az (2003), Campbell et al. (1986), Garnett (1992), and  
Oey \& Shields
(2000). The latter three are identical and are based  on Stasi\'nska`s (1980) models. For 
all of parametrizations studied, the average differences are less than $500$ K, except for that of 
Pilyugin et al. (2006), for which the average difference is of the order of $800$ K. Therefore, our $T(O^{+})$ value 
agrees with those obtained from other parametrizations. We also determined the $T(O^{+})$
from the RO2 ratio using the intensities of $\lambda7320+\lambda7330$ given by Kniazev 
et al. (2004). The determination was done using both Equations (2) and (3). In both cases, 
the differences between the $T(O^{+})$ from equation 1 (or from any of the parametrizations 
detailed above) and from equations 2 and 3 are about $3000$ K. We think that this is due to 
more deeper problems which are out of the scope of this paper and which might be studied in 
detail in a separate paper. In any case, the values of the $T(O^{+})$ used here in the oxygen
abundance determinations are not  worse than any other determined from parametrizations. 

The uncertainties in the oxygen abundances determined here were obtained 
from  the uncertainties in the line intensities reported by Kniazev et al. (2004). The 
error bar was made symmetric with the mean value taken as the nominal value. The average
uncertainty in the oxygen abundance is about $0.08$ ($\sigma$ = $0.03$). Only $9\%$ of 
the uncertainty values are larger than $2\sigma$. The differences between the uncertainties
determined by us and those reported by Kniazev et al. (2004) in their table 4 are less than 
$0.02$ for the  majority of the galaxies. 

The next thing to be done before comparing the standard method with the semiempirical methods is to test the internal consistency of the standard method itself. It
is important to know how accurate the values obtained by our code are compared with other
values obtained from other procedures. It is claimed that differences in the procedures
might result in important differences in the $T_e$ and, therefore, in the oxygen abundances.

In order to check this,  we compared our abundance data with those obtained from the 
formalism of (1) Kniazev et al. (2004), (2) 
Izotov et al. (2006), and (3) Aller (1984). The results are shown in Table 1. The first
row shows the average differences in the oxygen abundances between our values and those 
listed above. The second row shows the  percentages of those values which have differences 
larger than the average uncertainties. The percentages of those values with no differences
at all are shown in the third row. It can be seen that the largest differences are obtained 
with the expression by  Izotov et al. (2006), while the other two oxygen abundances are almost
identical to our values. This is in good agreement with Izotov et al. (2006), who claimed 
that the differences in the oxygen abundances are only $1\%$ when the original equations by
Aller (1984) are used in their derivation. 

\subsection{On the semiempirical methods}

As previously said, there are several methods based on the calibration of the strongest 
spectral lines to derive chemical abundances, usually oxygen abundances, when the auroral
transitions are not detectable. The most common are those based on the [OII]+[OIII]/H$\beta$ 
ratio, in particular, the $R_{23}$ calibrator proposed by Pagel et al. (1979) and the $P$ 
method. There are some other methods for determining the chemical abundances based on other 
lines: the S$_{23}$ (D\'{\i}az \& P\'erez-Montero 2000), S$_{234}$ (Oey \& Shields 2000), 
the [N\,{\sc ii}]/H$\alpha$ ratio (Denicol\'o et al. 2002), the $S_{23}$/$O_{23}$ parameter 
(P\'erez-Montero \& D\'{\i}az 2005, and the O$3$N$2$ by Pettini \& Pagel (2004). We did not 
study them because the lines needed are not available for the sample we used. 

The $R_{23}$ method is based on the behaviour between the [OII]+[OIII]/H$\beta$ ratio 
(the $R_{23}$) and the metallicity. The former increases while diminishing the latter 
because of both the enhanced heating and the diminished cooling (mainly due to O$^{++}$, 
O$^{+}$, and S$^{++}$). Therefore, at low metallicity the $R_{23}$ ratio increases with 
the oxygen abundances. At high metallicities, the greater cooling efficiency pushed 
more of the collisionally excited lines energy into the infrared, and [OII]+[OIII]/H$\beta$ 
increases with diminishing oxygen abundance. The validity of this method depends 
on the existence of a statistical relationship between the ionization temperature 
of the hottest stars ($T_{ion}$) and the oxygen abundances (Pagel et al. 1979). Therefore, 
[OII]+[OIII]/H$\beta$ 
is nearly invariant with respect to the geometrical factors (not with the ionization factor) 
but varies smoothly with the $T_{ion}$ (Olofsson 1997). 

In his study, McGaugh (1991) used the photionization code CLOUDY to reproduce detailed 
H\,{\sc ii} regions models. He used star clusters as ionizing source of the H\,{\sc ii} 
regions, and the metallicity of the ionizing stars is taken into account. The main 
caveat is that the clusters are zero-age and any evolutionary effect influencing the 
equivalent effective temperature is taken into account. Also, he considered the 
dependence on the ionization parameter of the ratio [OII]+[OIII]/H$\beta$. He found 
that such dependence varies with the metallicity so that the parametrization is not easy. In any case, a set of equations can be obtained for each of the branches 
in which the metallicity range is divided. The main caveat is that photoionization 
models have serious problems of uniqueness of fitting because there are many 
disposable parameters: the $T_{ion}$ of the star(s), the chemical composition, the 
clumpiness of the density, the geometrical factors, etc, (Pagel et al. 1979). All of 
them affect the size of the H\,{\sc ii} regions and, therefore, the state of ionization. 
Considering all the problems, the estimated accuracy of this calibration is about 
$0.15$ dex (Kewley \& Ellison 2008). Hereafter, the abundance determined with 
this calibrator will be called the $R_{23}$ abundance.

The next method we studied is the one introduced by Pilyugin (2000, 2001). He introduced 
a new parameter (the $P$ parameter) defined as the contribution of the radiation on the 
lines [O\,{\sc iii}]$\lambda,\lambda$4959,5007 and [O\,{\sc ii}]$\lambda,\lambda$3726,3729
to the total oxygen radiation and calibrated it for a sample of H\,{\sc ii} regions.  
Such a parameter corrects for the effect of the ionization parameter and takes into account 
the physical conditions  of the regions because it is an indicator of the hardness of the
ionizing radiation. Recently, a new sample was used to recalibrated the high-metallicity region (Pilyugin \& Thuan 2005). They used a large sample of data which 
includes the entire range of excitation. Hereafter, the abundance determined with this 
method will be called the $P$ abundance.

\section{Standard versus semiempirical methods}

In order to know which method yields abundance determination closer to those derived by 
the SM, the abundances determined with the SM and with each one of the 
semiempirical methods for all the galaxies in the sample were compared. As previously 
said, our sample has been observed, reduced, and analyzed in the same way. Therefore, 
the differences in the abundances between the different methods cannot be due to 
differences in the analysis.

In the calibration of the $R_{23}$ parameter three branches are differentiated: the 
low-metallicity branch (12+log(O/H) $< 8.1$), the high-metallicity branch (12+log(O/H) 
$> 8.4$), and the ``turnaround''  region ($8.1~ <$ 12+log(O/H) $< 8.4$; e.g.,  McGaugh 
1991). Pilyugin (2000) obtained a good correlation between his $P$ parameter and the 
oxygen abundances for 12+log(O/H) $< 7.95$ and 12+log(O/H) $> 8.2$. Afterwards, he 
divided his sample into two regions: the low metallicity (12+log(O/H) $< 8.2$) and 
the high metallicity one (12+log(O/H) $> 8.2$; Pilyugin 2001). Combining these two
partitions and the results of figure ~\ref{fig2}, we decided to  divide the total 
sample into three different regions: the low-metallicity (12+log(O/H) $< 7.95$), 
the high-metallicity branch (12+log(O/H) $> 8.2$), and the ``turnaround''  region 
($7.95~ <$ 12+log(O/H) $< 8.2$) in order to simplify the study. We have also taken into account the necessity of a detailed study of the ``turnaround'' 
region, which seems to be the most troublesome; e.g.,  Melbourne \& Salzer 
(2002) did not use the $R_{23}$ abundances for those galaxies of their sample located 
at this region due to their low confidences on the results. In order to do such 
study, a sufficient width in the ``turnaround''  region is needed, which will not 
be reachable if the traditional cut at $8.1$ and the cut of the 
high-metallicity region by Pilyugin at $8.2$ is considered. Instead, we used the first cut of the 
low-metallicity region by Pilyugin (2000) at $7.95$ as the lower limit of the 
``turnaround''  region. The main caveat might be that our regions are very small.
Our ``turnaround''  region is only $0.05$ dex smaller than the one 
defined by McGaugh (1991). Our high metallicity region ends up at $8.5$, mainly due 
to the requirement that the galaxies in the sample should have the 
oxygen auroral transition line detected. This line is not easily detected for metallicities
above $8.5$. The problem in the low-metallicity region is due to the original 
sample itself. The number of low-metallicity galaxies detected by SDSS is 
very small: only $200$ in front of several thousands of galaxies with metallicities
larger than $7.5$ were obtained in the later release (Thuan  2008). 
Therefore, the number of very low-metallicity galaxies in the first releases of 
SDSS is very small.  

With such limits, we can say that in our sample there is a total of $65$ galaxies 
in the high-metallicity region, $100$ in the low-metallicity one and $285$ in the 
``turnaround''  region, determined from the SM abundances. Although the abundance
range is narrow, just one order of magnitude, we think that there are enough 
galaxies in each of the three regions to obtain reliable conclusions. 
Finally, is has to be remarked that the abundances of all the galaxies in our sample 
are subsolar and any conclusion obtained in this investigation is valid only to 
this range of metallicity and should not be extrapolated outside it.

First, we  will focus on the distribution of the abundances of the 438 galaxies of 
our sample. The total abundance range is $7.6$-$8.5$. The histogram with the SM 
abundances distribution is shown in Fig ~\ref{fig2}a. The peak, corresponding to 
the most probable value, is equal to $8.1$.  Thirty-two percent of the galaxies of the total 
sample have this abundance. One hundred and ninty-six (43$\%$) galaxies are less metallic, and $111$ 
(25$\%$) are more metallic, than $8.1$. Therefore, it is an asymmetric curve. The 
distribution determined with the $R_{23}$ 
is shown in figure ~\ref{fig2}b. It is very similar to figure ~\ref{fig2}a: a large 
peak is at $8.1$ ($38\%$ of the total sample). The main difference is the narrow range 
in metallicity with no galaxies more metallic than $8.3$, indicating a large width 
of the curve. On the other hand, the 
distribution of the $P$ method is very different. It shows a bimodal distribution, 
with two peaks corresponding to the low and high metallicity regions. Such a 
distribution is not observed for any other method and is likely due to the two 
different equations used in the determination of the abundances for high ($>$$8.2$) 
and low ($<$ $8.2$) metallicity (Pilyugin \& Thuan 2005). The distribution 
shown in figure ~\ref{fig2}c differs  with the previous ones not only in the shape 
but also in the most common value. The low-metallicity peak is at $7.95$, covering 
two bins in metallicity from $7.9$ to $8.0$. Sixty-one percent of the galaxies have $P$ 
metallicities in this range. Concerning the high metallicity, the peak is at $8.3$ 
with a total of $48 \%$ of the galaxies having such values of the abundance.

The next step is to compare the abundances from the SM and the $R_{23}$ methods. This
is shown in figure ~\ref{fig3}. The solid line is the 1:1 line, where both methods 
give the same abundances, while the dashed line is the dispersion of the $R_{23}$ 
calibrator determined by us and without the typical uncertainty
for this method. The lines are not parallel because of the differences in the dispersion 
in the three metallicity regions. Due to the small metallicity range, the three abundance 
branches are shown in one single plot, with dashed vertical lines dividing them. As 
seen in figure ~\ref{fig2}b, one effect of the 
$R_{23}$ calibrator is to narrow the metallicity range. This can be seen also in 
figure ~\ref{fig3}, where only one low-metallicity galaxy has a $R_{23}$ value lower 
than the SM ones. Moreover, only one high-metallicity galaxy has a $R_{23}$ abundance 
higher than the SM one. In contrast, $25$ ($13$) have higher(lower) metallicity 
than the value determined with the SM for low- and high-metallicity regions,
when considering the uncertainties of the SM data. In order to get the goodness of 
the $R_{23}$ method, the dispersion can be evaluated. It gets values of $0.15$, $0.19$, 
and $0.09$ for the high, low, and ``turnaround'' regions (See Table 2). Finally,
the percentage of data points which do not fit the 1:1$\pm$dispersion line are  
$20 \%$, $25 \%$, and $9 \%$ for the high, low, and ``turnaround'' region, respectively. 
Then, the values of the $R_{23}$ method at the ``turnaround'' region are confident enough, 
in contrast to what Melbourne et al. (2004) claimed.

Figure ~\ref{fig4} shows the same plot but for the $P$ method. The abundances 
have been determined with the low-metallicity equation (Pilyugin 2000) for the 
galaxies in the turnaround and the low-metallicity regions and the new calibration
by Pilyugin \& Thuan (2005) is used for the high-metallicity region. The first thing 
to notice is that the distribution is not smooth now, in the sense that there is 
a discontinuity at $8.2$. Moreover, most of the data points in the ``turnaround'' 
region do not fit the 1:1$\pm$dispersion line, but they give smaller metallicities 
than the SM ones. This might be likely because the abundance determination in the 
``turnaround'' region was done using the low-metallicity equation, following Pilyugin 
(2001). But previously, he used this same equation only for galaxies with abundances 
smaller than $7.95$. Therefore, we can concluded that the equation for the low-metallicity branch is not entustred for the ``turnaround'' region. This also 
can be checked using the dispersions. The value for the `turnaround'' region is the 
highest one ($0.14$, compared with $0.09$ and $0.11$ for the high- and low-metallicity 
regions, respectively) but still there are $82\%$ of the data points in this region which do 
not fit the 1:1$\pm$dispersion line.

In Table 2 we have summarized the dispersion and the fitting percentages for the 
two calibrations for each one of the metallicity regions. The first thing to notice 
is that the dispersion is not constant with metallicity for any of the semiempirical 
methods studied here. There is a dependence on the metallicity so a single value 
cannot be used for the whole range. The best numbers for the low and high regions 
are given by the $P$ method, but for the ''turnaround`` region the best numbers are 
those obtained by the $R_{23}$ method, due to the problems of the $P$ method mentioned above. 
Another reason might be because the $R_{23}$ abundances are clustered around $8.0$, 
as can be seen in figure ~\ref{fig2}a.
 
With these values it can be concluded that the semiempirical methods provide 
values for the metallicity closer to those of the SM ($\sigma < 0.1$) for six 
out of ten galaxies studied here. Moreover, the $P$ method gives the closest values 
to the SM abundances for those data points with abundances smaller than $7.9$ and larger 
than $8.2$.

\subsection{On the dependence of the dispersion}

One of the most interesting features found in this investigation is the fact 
that the dispersion varies with the metallicity. In order to understand it,
we study some parameters to see on which of them the dispersion 
depends. 

In figure ~\ref{fig5}, the residuals between the SM abundance and the $R_{23}$(a) 
and the $P$(b) abundances are shown. The uncertainties of the SM method are 
shown as dotted lines. The first thing to notice is the discontinuity 
in the residuals for metallicities larger than $8.2$ for the $P$ method. This is 
probably a consequence of the two different equations used in the determination 
of the abundances with this method. Secondly, there is a correlation between the 
metallicity and the residuals, it being positive for values larger than $8.0$ 
and negative for values lower than $8.0$. Such correlation is present for both 
methods, the strongest being that for the $R_{23}$ with residuals of up to $0.6$, 
while the largest residuals for the $P$ method is $0.43$. In spite of the 
relationship, only $13\%$ and $12\%$ of the galaxies have residuals larger than 
$0.2$ for the $R_{23}$ and the $P$ methods, respectively. This might indicate the 
low sensitivity of the semiempirical methods to the real value of the metallicity, 
giving values of $8.0 \pm 0.2$ for most of the data.

Our figure ~\ref{fig5} can be compared with Figures 5 and 6 in P\'erez-Montero \& 
D\'{\i}az (2005). They used a different metallicity range than that considered 
in the present investigation. Therefore, the comparison is restricted to the common metallicity range. The only difference is that they considered the differences between 
the semiempirical methods and the SM method. Therefore, they got negative residuals 
at larger metallicities. They do not obtain any gap for values larger than $8.2$ 
for the $P$ method, but they present the low and high metallicity branches in two 
different figures and so the gap might be difficult to see. Their residuals are 
larger than our values but the correlation is similar. They also found out that 
the larger values of the residuals are for abundances of around $7.6$, diminishing 
for lower metallicities. As we do not have in our sample very low metallicity galaxies 
(12+log(O/H) $<$ 7.6) we cannot investigate this point with our more homogeneous sample.

We can study whether there is any other dependence on the residuals. In particular we are 
interested in the possible relation between the residuals and the ionization parameter, 
([O\,{\sc ii}]/[O\,{\sc iii}]), mainly because both methods used here claim that 
such a parameter is considered in the calibrations. The relationship between the 
([O\,{\sc ii}]/[O\,{\sc iii}]) ratio and the residuals of the $R_{23}$ is shown 
in figure ~\ref{fig6}a, while figure ~\ref{fig6}b shows the same relationship 
for the $P$ abundance. The behaviour for both, the $R_{23}$ and the $P$ residuals 
with the ([O\,{\sc ii}]/[O\,{\sc iii}]) ratio are very similar: for those galaxies 
with log([O\,{\sc ii}]/[O\,{\sc iii}]) $>$ 1.0 (high-ionization regions) the 
residuals get more negative, indicating that the abundances determined with the 
semiempirical mehtods are larger than the SM ones. 

Another interesting parameter will be the equivalent width of H$\beta$, EW(H$\beta$).
This is a good age indicator (see section 3.2). The plot between the residuals and 
the EW(H$\beta$) is shown in figure ~\ref{fig7}. In this case, there is no clear 
trend, apart from the fact that for those galaxies with EW(H$\beta$) smaller than 
$30$ and lager than $150$, the residuals are more negative for the $R_{23}$ method. 
The trend is similar for the $P$ method but the dispersion is larger. 

In conclusion, it can be said that the residuals depend mainly on the abundances 
and, probably on the ionization parameter.

\subsection{The influence of the age of the regions}

P\'erez-Montero \& D\'{\i}az (2005) from their study of the semiempirical methods 
concluded that there are not many ways to improve the $R_{23}$ and the $P$ methods 
because of their dependence on the ionization parameters and ionization temperatures. 
These parameters change as the H\,{\sc ii} evolve in a way that is not easy to 
parametrize. 

In order to check the influence of the difference of the age of the star formation 
bursts we can divide the sample considering their morphological types. It is well 
known that blue compact galaxies (BCG) are experiencing right now an intense burst of star formation, while 
the events of star formation in both late spirals (Sm) and irregular (Im) galaxies are less intense. In consequence, 
the ionization parameter and temperatures might be different for these three 
subsamples of galaxies, and therefore, the $R_{23}$ and $P$ abundances. The main 
caveat is that the differences between these morphological types are very subtle, 
especially  between the BCG and the Im. It is considered that BCG have younger 
bursts of star formation, but their EW(H$\beta$) range is between $14$ and $280$, 
according to table 2 in Kniazev et al. (2004). 
Such values might indicates that the ages of the present burst in the BCG are 
between  $4$ and $10$ Myr for a Salpeter IMF and an upper limit mass of $120~M\odot$,
following the Copetti et al. (1986) models. Moreover, all the Sm and $18$ of the Im 
have EW(H$\beta$) smaller than $50$. Therefore, it can be concluded that not all 
the BCG have truly young H\,{\sc ii} regions but some of them are as old as the bursts 
in Sm and Im. In contrast, all the Sm have old bursts. In consequence, the 
morphological type is not a good approach to study the influence of the age.

Another approach to the problem of the age/evolution of the regions is using a 
parameter to discriminate between truly ``old'' and ``young'' H\,{\sc ii} regions. 
As previously said, equivalent width of H$\beta$ has been considered as a good 
age indicator (Copetti et al. 1986). Therefore, we can explore such parameter 
in our sample in order to see the age effect. The first thing to notice is that 
269 out of the 438 galaxies have values of the EW(H$\beta$) smaller than $50$, 
independently of the morphological type. This EW(H$\beta$) implies an age of the 
burst of at least $8$ Myr, and it is quite independent of the parameters of 
the models (Copetti et al.  1986). Another interesting fact is that only $19$ 
galaxies with EW(H$\beta$) larger than $50$ have low abundances, while $69$ 
of the galaxies with EW(H$\beta$) smaller than $50$ show a low metal content. 

Finally, we can explore the goodness of the $R_{23}$ and the $P$ method using 
this parameter. Figure ~\ref{fig8} shows the SM versus $R_{23}$ (Figure 8a) and the SM versus $P$ (Figure 8b) abundances, but now the galaxies with EW(H$\beta$) larger than $50$ 
are plotted as triangles. It is interesting to see that the majority of the 
galaxies which do not fit the 1:1$\pm$dispersion line have EW(H$\beta$) $>$ 
$50$ ($53$ out of $70$) with the $R_{23}$. The plot of the $P$ method is more 
complex: at low metallicity all of the galaxies outside the 1:1$\pm$dispersion 
line have EW(H$\beta$) $>$ $50$, while at the ``turnaround'' region it is the
opposite. At high metallicity, only half of the sample outside the 
1:1$\pm$dispersion line have large values of the EW(H$\beta$). The conclusion 
emerging from these plots is that those galaxies with younger bursts of star 
formation might have large discrepancies between the SM and the $R_{23}$ (or 
$P$) abundances.

\section{Comparison with other investigations}

The comparison of the different semiempirical methods against the SM has been done  
before (e.g., P\'erez-Montero \& D\,{\'i}az 2005 and references therein). All of them 
follow the same procedure. In order to create a large sample the authors gathered data from 
the literature. These data were very heterogeneous in many ways: they were observed 
with different apertures, were analyzed by different persons with different criteria 
(see A.M. Hidalgo-G\'amez 2009, in preparation, for a study of the variation in the SM 
abundances due to these effects), etc. Therefore, the results cannot be completely 
reliable. The dispersion might account for the differences in the acquisition/reduction/analysis 
procedure. In this sense, the sample studied here  is completely homogeneous. It 
is large enough to obtain conclusive results. Another advantage of the results presented 
in Section 3 is that the range in metallicity has been divided into the three typical 
branches and therefore they can be studied in more detail. Here we are going to compare 
our results with some of the most recent investigations. The only caveat is that the
results are restricted to the abundance range studied here and should not be extrapolated
outside it.

Lee et al. (2003) made a comparison between the SM and the $R_{23}$ calibrator concluding 
that most of the data was consistent with the 1:1 $\pm 0.2$ dex line. These uncertainties 
are too relaxed, as can be seen in Table 2. But when the uncertainty is only $0.1$, 
more than half of their data are located outside the region, consistent with the result 
presented here. Moreover, they did not make the separation into the three branches. 
When such a division is done, $38\%$, $70\%$, and $50\%$ of the galaxies do not lay in the 
1:1$\pm$0.1 dex. These values are again very similar to those presented here. 

Kennicutt et al. (2003) made a comparison between the SM abundance and the $R_{23}$ 
one for H\,{\sc ii} regions with metallicities larger than $8.0$. Their sample is 
much smaller than the one presented here. Their $R_{23}$ metallicities are higher 
than their SM abundances. This is the opposite of what we found here. Several reasons 
might play a role: the differences in aperture among the galaxies in their sample, 
the morphological differences, or a problem with the photoionization models used in 
the $R_{23}$ calibration from Kewley \& Dopita (2002), as they discuss. Surprisingly, 
the dispersions are very similar to those in figure~\ref{fig3}. They also compared 
the $P$ metallicity (old calibration) with the SM abundances. For high metallicity, 
their $P$ value is higher than or equal to the SM value, but at metallicity $\approx$ 
$8.0$ the $P$ abundance is lower than the SM abundance. Although we did not study 
the old calibration of the $P$ method, a similar trend is found in Figure 
~\ref{fig4}, using the new calibration. Also, the dispersion at high metallicity is 
similar in both investigations in spite of  the differences in the sample size.

P\'erez-Montero \& D\'{\i}az (2005) conducted a study of the different nebular calibrator and 
photoionization models. They used several calibrations for the $R_{23}$ abundance but we focus on their results with the McGaugh (1991) calibration because this is the 
calibration used in the present investigation. They used a heterogeneous sample with 
data from different sources in the literature. They found that at high abundance the 
$R_{23}$ abundances are larger than those with the SM with a dispersion of $\approx$ 
$0.7$. For low metallicities, the results are similar to those shown in figure ~\ref{fig3}, 
but with larger dispersion. They also compared the $P$ abundances; again, they found 
similar results to those found in the present investigation, but with larger dispersion.  

Recently, Shi et al. (2006) presented a comparison between the SM and several 
semiempirical methods similar to the one presented here. Their main advantage is 
the number of data points: a total of $4222$. But their main drawback  is that they 
did not use the SM method for $3997$ galaxies, where the [OIII] $\lambda$4363 line 
was not detected. Instead, they used equation 11 in Pilyugin (2001) to determine the 
$T_e$ and afterwards, the relations by Garnett (1992) and Pagel et al. (1992) to 
determine the rest of the parameters.  Therefore, their sample is not self-consistent 
because they mixed different methods in the determination of the oxygen abundances of 
a single galaxy. So, the oxygen abundances of their sample II cannot be used for 
comparing the different methods. The main advantage is that their sample is not biased
to galaxies with the oxygen line [O\,{\sc iii}]$\lambda$4363. In any case, and considering 
only their sample I of $225$ galaxies, their results are very similar to those 
presented here for the R${23}$ and the $P$ calibrator. 

Finally, Liang et al. (2006) studied the different calibrators with a very large 
sample (over $40,000$ galaxies). Again, they have to use the $R_{23}$ calibrator in 
order to determined the abundances for most of the galaxies of their sample. The small 
metallicity range in common between their sample and the study presented here makes 
any comparison meaningless. 

\section{Discussion}

Melbourne et al. (2004) made a comparison of the abundances determined with the 
SM and the $P$ calibrator. They found that the most deviating galaxies of their 
sample of low-metallicity galaxies correspond to those with the lower excitation 
indicator ([OIII]/[OII], as defined by them) and high electronic temperature. 
Figures ~\ref{fig9} and ~\ref{fig6} show the $T_e$ and the [OII]/[OIII] versus the 
residuals between the SM and the $P$ methods, respectively. For low-metallicity 
galaxies we obtained the same results as Melbourne et al. (2004): the most 
deviating galaxies, defined as those with $SM-P > 0.2$, are those with $T_e$ 
larger than $14,000$ K . In the case of those galaxies with high metallicity 
the most deviating are those with $T_e$ smaller than $10,000$ K (11,000 K in 
the $P$ method). The most deviating galaxies in the ``turnaround'' region have 
low $T_e$, just the opposite of that in the low-metallicity region. This behaviour 
might be due to the correlation shown in figure ~\ref{fig1} for galaxies with 
$T_e$ larger than $14,000$ K, because larger abundances mean lower $T_e$ and 
lower [OII]/[OIII], but it is not related to any physical properties of the 
galaxies.  

In any metallicity range the cutoff in the ionization parameter is not that 
clear. The only trend is that those galaxies with a low ionization parameter 
have a semiempirical oxygen abundances that are similar to the SM value ($\pm$ 
0.2). The discrepancy with the SM abundances is larger for those galaxies with 
larger values of the ionization parameter.

\section{Conclusions}

We have made a detailed comparison of the abundances obtained with the so-called 
standard method (SM) and some of the semiempirical methods used: $R_{23}$ 
(McGaugh's calibration) and $P$ (Pilyugin \&  Thuan 2005). Our 
main interest is to obtain the closest abundances to the SM abundances when the standard 
method cannot be applied. In order to do that we used a large sample of late-type 
galaxies observed by SDSS and reduced by Kniazev et al. (2004). For all of the 
galaxies in the sample the oxygen forbidden line [O\,{\sc iii}]$\lambda$4363 is 
detected, and therefore the SM abundances can be determined. The main advantage of 
this sample is that it is large enough for obtaining conclusive results and very 
homogeneous (being observed by only one facility and reduced and analyzed in the 
same way for all the galaxies); therefore, the systematic errors will affect all 
the data in the same way. This is not the usual procedure in the literature, 
where data, gathered from different sources, are obtained from different facilities, 
and analyzed and reduced by different researchers. A.M. Hidalgo-G\'amez (2009, in preparation) 
found that this procedure might be a large source of uncertainties in the 
abundances determination. One of the 
disadvantages of the sample is that the use of only those galaxies where the 
auroral oxygen line was detected restrict the metallicity range, especially at 
high metallicity. Therefore, and keeping this last statement in mind, the 
conclusions obtained in this investigation are restricted to sample with 
similar metallicity range and characteristics.

The main conclusions could be listed as follows.

\begin{figure}
\centering
\includegraphics[width=10cm]{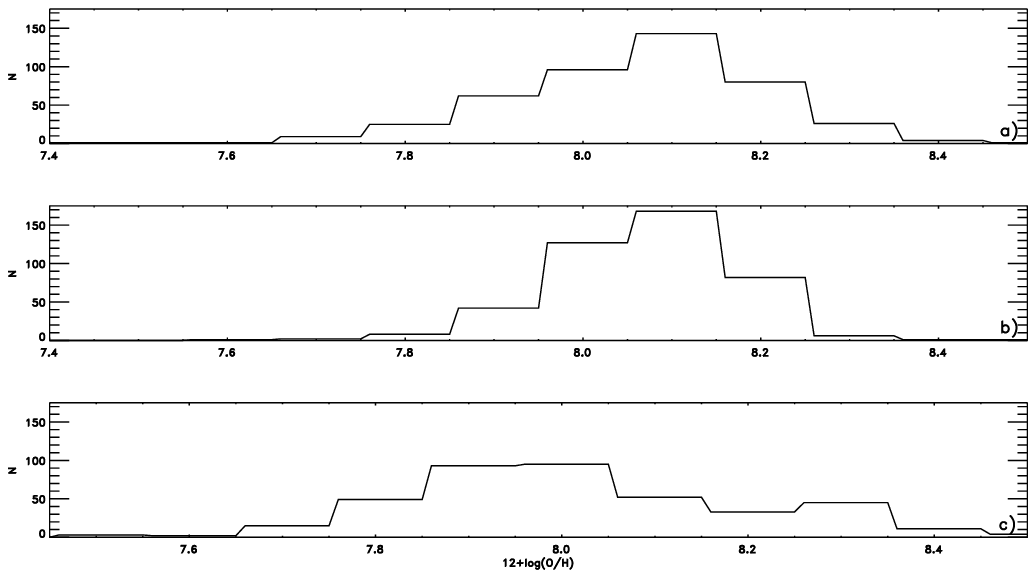}
\caption{The metallicity distribution of the total of $438$ galaxies studied in the 
present investigation. The metallicity has been determined with the standard method 
(a), the $R_{23}$ method (b), and the 2005 calibration of the $P$ method (c). The 
distribution of the two first methods is very similar, while the $P$ 
method gives a bimodal distribution due to the two different equations used for the 
metallicity determination.}
\label{fig2}
\end{figure}

\begin{figure}
\centering
\includegraphics[width=10cm]{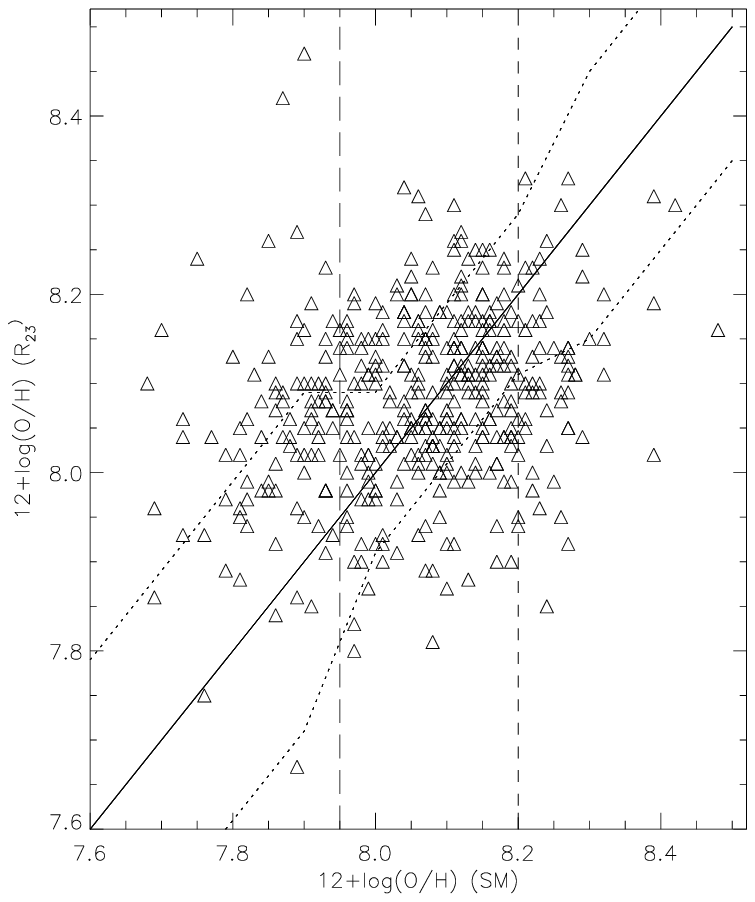}
\caption{
The metallicity determined with the SM method vs. the $R_{23}$ method for the 
438 galaxies in the sample. The solid line is the 1:1 line while the dashed ones are 
the dispersion. As the dispersion depends on the metallicity, they are not parallel. }
\label{fig3}
\end{figure}

\begin{figure}
\centering
\includegraphics[width=10cm]{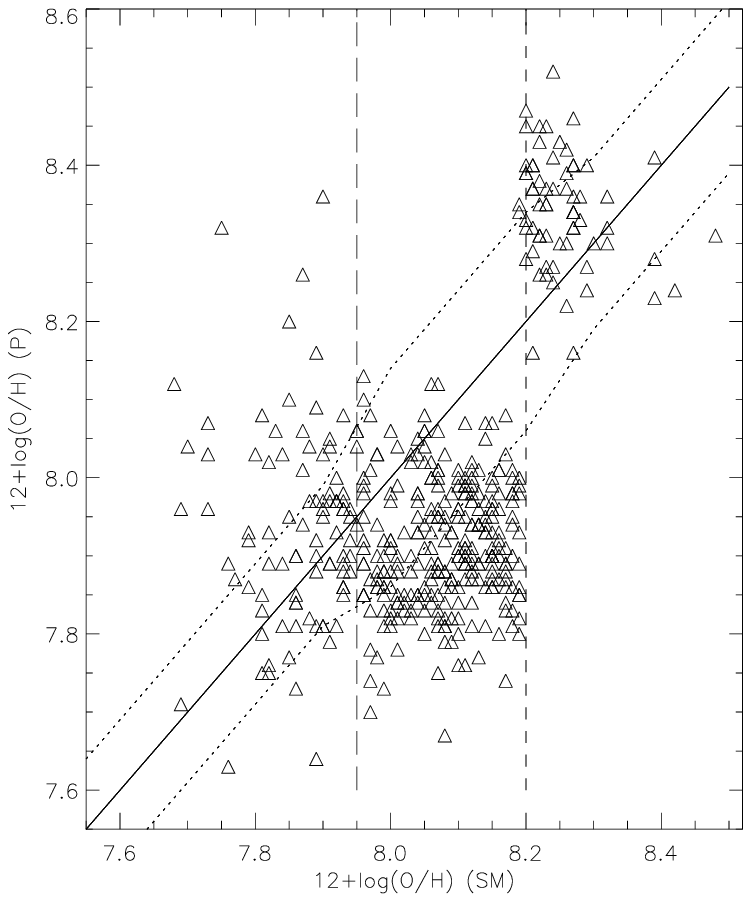}
\caption{The metallicity determined with the SM method vs. the $P$ method  
for the 438 galaxies in the sample. Symbols and lines as in figure ~\ref{fig3}. }
\label{fig4}
\end{figure}

\begin{figure}
\centering
\includegraphics[width=10cm]{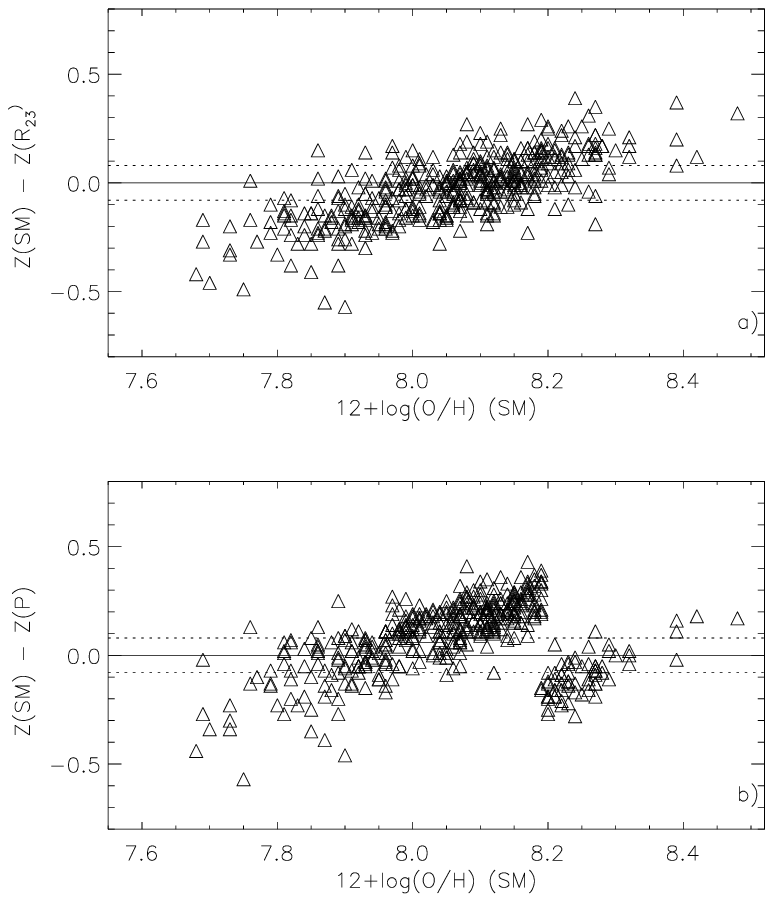}
\caption{Residuals between the SM and the $R_{23}$(a) and 
the $P$(b) methods vs. the SM abundances for the 438 galaxies in the sample. The solid 
line at $0.0$ indicates that both methods give the same value of the abundance, 
while the dashed lines are the uncertainties of the SM method.   }
\label{fig5}
\end{figure}

\begin{figure}
\centering
\includegraphics[width=10cm]{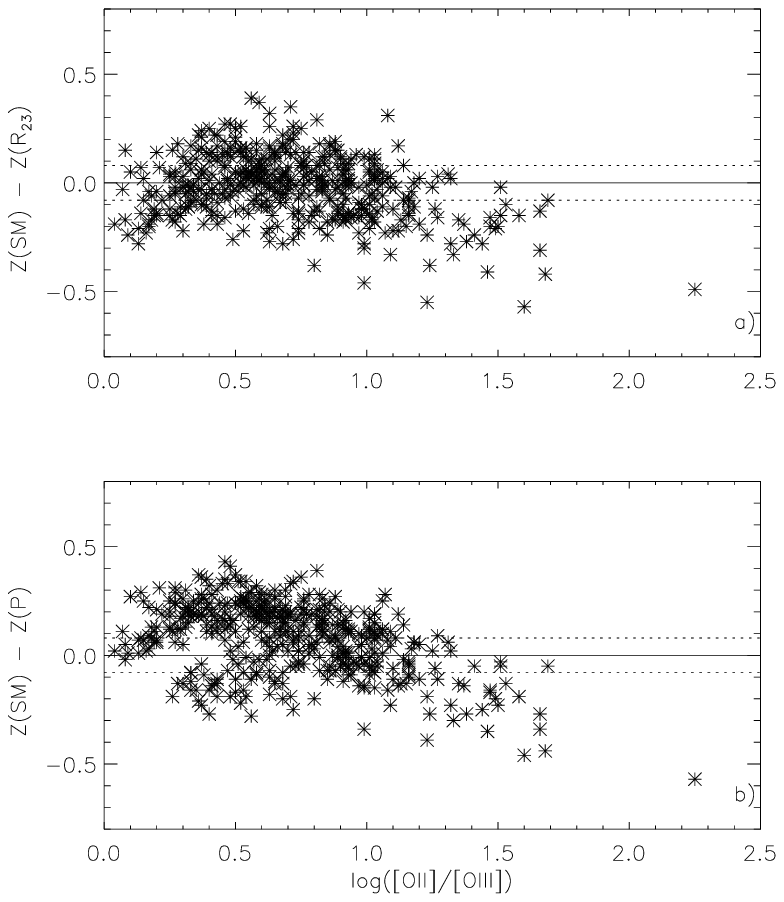}
\caption{Residuals between the SM and the $R_{23}$ (a) and the $P$ (b) methods vs. the 
log[O\,{\sc ii}]/[O\,{\sc iii}] ratio. Lines as in figure ~\ref{fig5}. }
\label{fig6}
\end{figure}

\begin{figure}
\centering
\includegraphics[width=10cm]{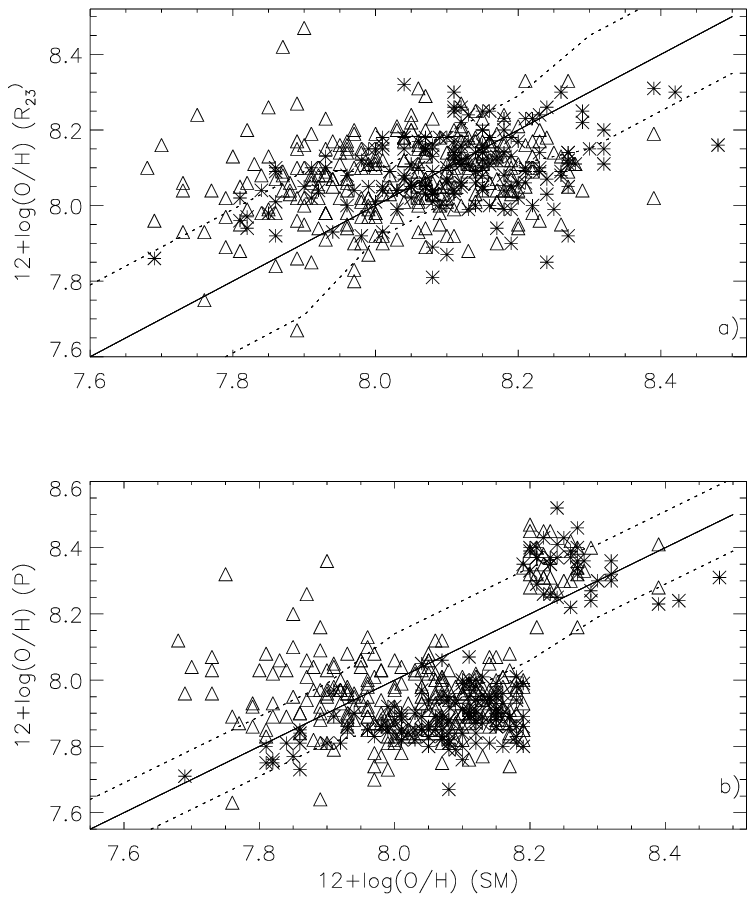}
\caption{The metallicity determined with the SM method vs. the $R_{23}$(a) and the $P$(b)
methods for those galaxies with EW(H$\beta$) $>$ $50$ (triangles) and smaller than 
$50$ (crosses). Lines as in figure ~\ref{fig3}.}
\label{fig7}
\end{figure}

\begin{figure}
\centering
\includegraphics[width=8cm]{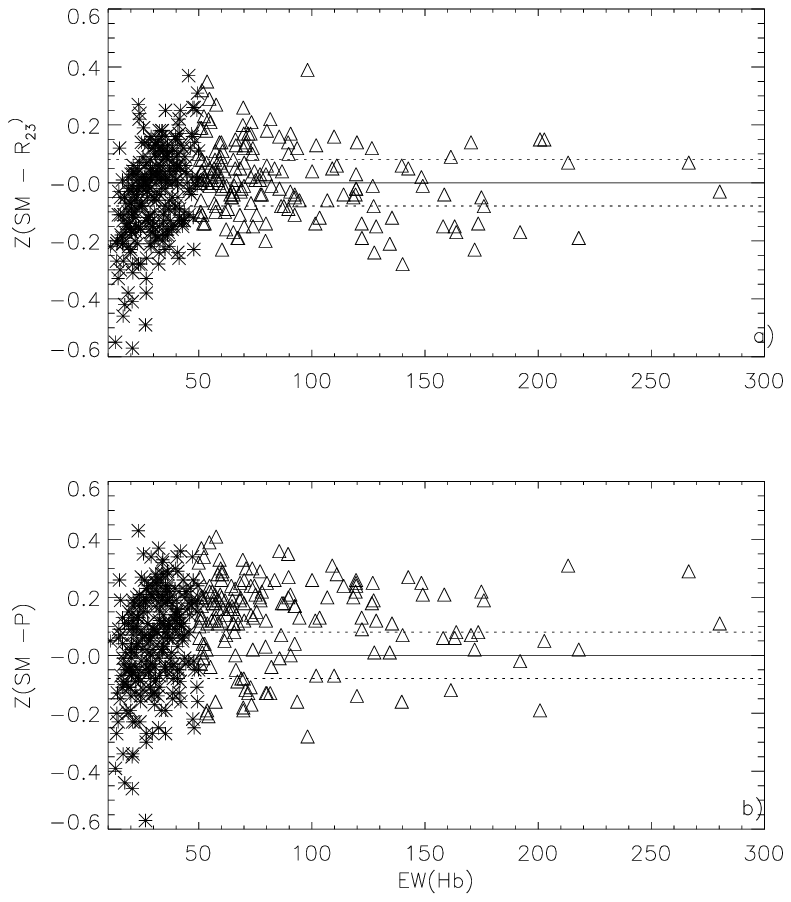}
\caption{Residuals between the SM and the $R_{23}$ (a) and the $P$ (b) methods vs. the 
equivalent width of H$\beta$. Triangles represent those galaxies with EW(H$\beta$) larger than $50$ and crosses those galaxies with EW(H$\beta$) smaller than $50$. Lines as in figure ~\ref{fig6}.}
\label{fig8}
\end{figure}

\begin{figure}
\centering 
\includegraphics[width=8cm]{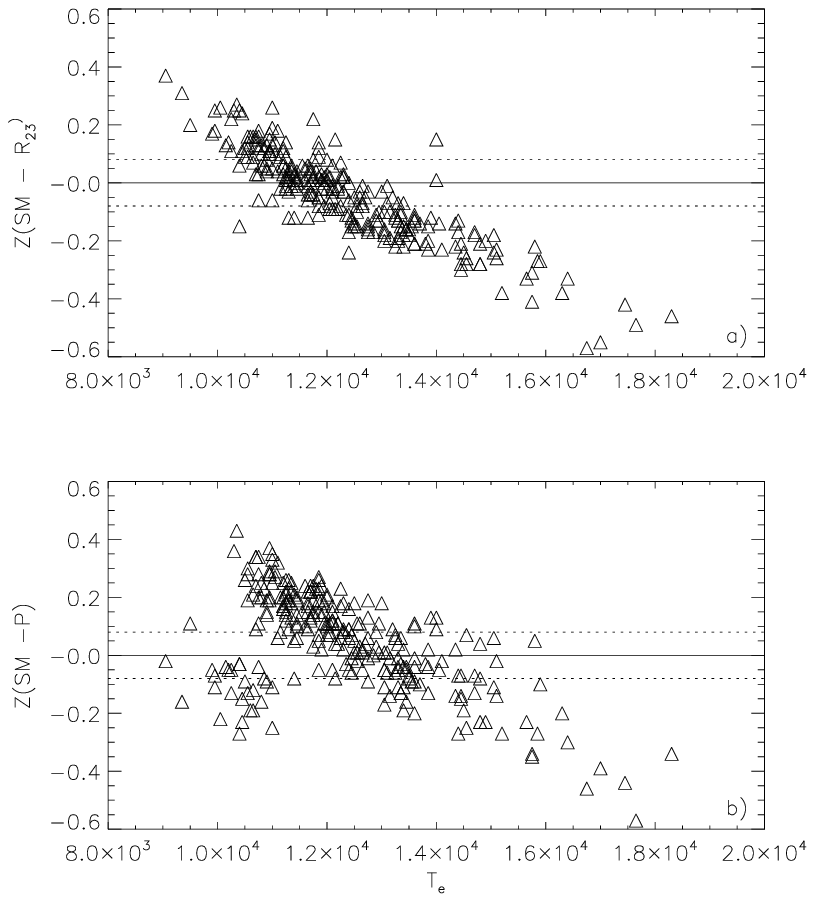}
\caption{Residuals between the SM and the $R_{23}$ (a) and the $P$ (b) methods vs. the 
electron temperature. Half of the galaxies with $T_e$ larger than $14,000$ K present 
a deviation larger than $0.2$. Lines as in figure ~\ref{fig6}.}
\label{fig9}
\end{figure}

1- The distribution of the metallicity of the galaxies is a single gaussian for both 
the SM and the $R_{23}$ method. The main differences between then is the smaller 
width of the gaussian for the $R_{23}$ abundances. This clustering of the $R_{23}$ 
abundances around a single value ($\approx 8.05$) is clearly visible at figures 
~\ref{fig2}b and ~\ref{fig3}. Such behaviour might be real or due to the sample 
itself, which is restricted to an ``special'' type of galaxies (those 
with the auroral oxygen line detected). Moreover, as the number of this type of galaxies is a tiny fraction of the total SDSS catalog, SDSS imposes important bias to any sample. In order 
to find out if the clustering is real, a similar study has to be done with other 
large sample obtained from choosing the galaxies observed without any ``a priori''
assumption.

2- The $P$ calibration gives the lowest dispersion in both the high- and the low-metallicity regions, while the $R_{23}$ gives the best results in the ``turnaround'' 
region. 

3- For both methods, there is a dependence of the dispersion with the metallicity 
for the low- and high-metallicity regions, while they are very insensitive to the 
metallicity in the ``turnaround'' region. 

4- There is a dependence of the residuals with the abundances and the $T_e$ for 
both methods. For galaxies with log[O\,{\sc ii}]/O\,{\sc iii}]$>$ 1, the residuals 
are larger. 

5- Finally, those galaxies with larger values of the equivalent width of H$\beta$ 
seem to have very different abundances than the SM abundances, especially at 
low metallicity, but there is no clear trend between the residuals and the EW(H$\beta$).

Finally, we might note that the results obtained here are similar to those found 
in other investigations, but due to the large and homogeneous sample used here, the 
results are more robust and the dispersion lower than in previous investigations.

\begin{acknowledgments}

The authors thank J. G\'onzalez and C. Morisset for many interesting suggestions 
which improved the paper and to J. Brenan for a carefully reading of the manuscript. 
The referee, Angeles D\,{\'i}az, is thanked for many interesting comments which have
improved the manuscript. A.M.H-G. thanks J.M. V\,{\'i}lchez for very 
interesting discussions. 
This investigation was supported by  DGAPA-UNAM grant IN114107 and CONACyT CB-2006. 

\end{acknowledgments}

\clearpage
\clearpage

\clearpage

\clearpage

\clearpage

\clearpage

\clearpage

\clearpage

\clearpage

\clearpage

\begin{table}
\caption[]{Differences Between the Oxygen Abundance Values determined with the code 
used in the present investigation and those of Kniazev et al. (2004) (column 1), 
Izotov et al. (2006) (column 2) and Aller (1984) (column 3). The second row shows 
the percentage of abundances values with differences smaller than the SM uncertainties
while the percentage of identical abundance values are shown in row 3.}
\vspace{0.05cm}
\begin{center}
\begin{tabular}{c c c }
\hline
{$\Delta$Z$_K$}  & {$\Delta$Z$_I$} & {$\Delta$Z$_A$} \\
\hline 
0.012          & 0.15         & 0.02     \\
3.4\%          & 70\%         & 5\%      \\
27\%           & 0\%          & 20\%      \\
\hline
\end{tabular}
\end{center}
\end{table}

\begin{table}
\caption[]{ Dispersion (top line) and fitting percentages (bottom line) 
for each of the metallicity regions given by the semiempirical methods 
studied here. The first column lists the regions, while the dispersion and 
percentages of the $R_{23}$ and P$_n$ are given in columns 2, 3 and 4. } 
\vspace{0.05cm}
\begin{center}
\begin{tabular}{c c c}
\hline
{Metallicity}  & {$R_{23}$}  & {P$_n$}\\
\hline 
high & 0.15           & 0.09 \\
     & 59\%       & 55\% \\
\hline
turnaround & 0.09      & 0.14 \\
           & 64\% & 62\% \\ 
\hline 
low  &  0.19           & 0.11 \\ 
     &  62\%      & 68\%\\
\hline
\end{tabular}
\end{center}
\end{table}

\end{document}